\documentclass[epj]{svjour}
\usepackage{epsfig,array}
\usepackage{multirow}
\usepackage{graphicx}
\newcommand{\be}{\begin{eqnarray}}
\newcommand{\ee}{\end{eqnarray}}

\newcommand{\nn}{\nonumber}
\newcommand{\bea}{\begin{eqnarray}}
\newcommand{\eea}{\end{eqnarray}}

\newcommand{\beq}{\begin{equation}}
\newcommand{\eeq}{\end{equation}}

\def\fun#1#2{\lower3.6pt\vbox{\baselineskip0pt\lineskip.9pt
\ialign{$\mathsurround=0pt#1\hfil##\hfil$\crcr#2\crcr\sim\crcr}}}

\begin{document}

\title{Photoproduction of $\eta$ mesons off neutrons from a deuteron target}

\author{
A.V.~Anisovich \inst{1,2} \and I. Jaegle \inst{3} \and E.~Klempt
\inst{1} \and B.~Krusche \inst{3} \and V.A.~Nikonov \inst{1,2} \and
A.V.~Sarantsev \inst{1,2} \and U.~Thoma \inst{1} }

\institute{Helmholtz--Institut f\"ur Strahlen-- und Kernphysik der
 Universit\"at Bonn, Nu\ss allee 14-16, 53115 Bonn, Germany \and Petersburg
Nuclear Physics Institute, Gatchina, 188300 Russia \and Institut
f\"ur Physik der Universit\"at Basel, Klingelbergstrasse 82, CH-4056
Basel, Switzerland }

\date{Received: \today / }

\abstract{A formalism is developed for the partial wave analysis of
data on meson photoproduction off deuterons and applied to
photoproduction of $\eta$ and $\pi^0$ mesons.  Different
interpretations of a dip-bump structure of the $\eta$
photoproduction cross section in the 1670\,MeV region are presented
and discussed. Helicity amplitudes for two low-mass $S_{11}$ states
are determined.}

\PACS{
     {11.80.Et}{Partial-wave analysis} \and
     {13.30.-a}{Decays of baryons} \and
     {13.40.-f }{Electromagnetic processes and properties} \and
     {14.20.Gk}{Baryon resonances with S=0} }
\authorrunning{A.V.~Anisovich {\it et al.}}
\titlerunning{ Photoproduction of $\eta$ meson off neutron}
\mail{klempt@hiskp.uni-bonn.de}

\maketitle

\section{Introduction}

The cross section for photoproduction of $\eta$ mesons off protons
is dominated by the $N(1535)S_{11}$ resonance, other resonances make
only minor contributions. In particular the $N(1650)S_{11}$
resonance, which one would naively expect to contribute to $\eta$
photoproduction in a similar strength as $N(1535)S_{11}$, is hardly
visible in the total cross section. There are two explanations why
$N(1650)S_{11}$ is so much suppressed compared to $N(1535)S_{11}$:
the $N(1650)S_{11}$ $\to N\eta$ decay branching fraction is much
smaller than that for $N(1535)S_{11}$ decays, and the
$N(1650)S_{11}$ photoproduction cross section seems to be suppressed
compared to $N(1535)S_{11}$ photoproduction.

The large $N(1535)S_{11}\to N\eta$ coupling found different
interpretations by Isgur and Karl \cite{Isgur:1977ef}, by Weise and
collaborators \cite{Kaiser:1995cy} and by Glozman and Riska
\cite{Glozman:1996tb}. In \cite{Isgur:1977ef}, the two quark-model
$\rm S_{11}$ states with $s=1/2$ and $s=3/2$, respectively, have
appreciable mixing (with a mixing angle of $-31^{\circ}$). A
phenomenological fit to baryon decays had given precisely this value
\cite{Hey:1974nc}. For this mixing angle, $N(1650)S_{11}$ decouples
from $N\eta$ decays while $N(1535)S_{11}$ has a strong coupling to
$N\eta$. In \cite{Kaiser:1995cy}, $N\eta$ and $\Sigma K$
photoproduction were described by the dynamics of the coupled
$\rm\Sigma K-p\eta$-system; no genuine 3-quark resonance was
required in their model. In \cite{Glozman:1996tb}, the
$N(1535)S_{11}$ is a conventional 3-quark state; one-pion exchange
was assumed to make an essential contribution to quark-quark
interactions. Clustering of the baryonic wave functions into quarks
and diquarks then led to the strong selectivity of the
$N(1535)S_{11}\to N\eta$ coupling.

The Moorhouse rule \cite{Moorhouse:1966jn} gives a second reason for
the small $\eta$ photoproduction cross section off protons. This
selection rule forbids photo-excitations off protons of spin 3/2
members of the lowest-mass [$70,1^-$] super-multiplet of nucleon
resonances. The Moorhouse rule forbids only transitions to the
[$70,^48$] component of the $\rm N(1650)S_{11}$. Burkert {\it et
al.} \cite{Burkert:2002zz} analyzed the existing photo- and
electroproduction  data and extracted transition amplitudes for
transitions from the nucleon ground state to the [$70,1^-$]
super-multiplet. The - scarce - data on electro-production off
neutrons were not used; due to their large errors, the data at the
photo-point hardly constrained their analysis.

Recently, photoproduction of $\eta$ mesons off neutrons has
attracted additional interest. A narrow structure at 1.67 GeV was
observed which was not easily understood in terms of known nucleon
excitations. It was first reported by the GRAAL collaboration at
NSTAR2004 \cite{Kuznetsov:2004gy} and interpreted as narrow
resonance by part of the authors
\cite{Kuznetsov:2006kt,Kuznetsov:2007gr}. The bump-like structure in
the $n\eta$ invariant mass distribution is not seen in the cross
section on the proton even though a reanalysis of the GRAAL data
\cite{Kuznetsov:2007dy} indicated the possibility of a bump
structure at 1.69\,GeV also for proton data. This structure was
suggested to signal the existence of a relatively narrow ($M\approx
1.68$\,GeV, $\Gamma\leq 30$\,MeV) baryon state. In particular the
possibility that the state is the non-strange member of an
anti-decuplet of pentaquarks
\cite{Diakonov:1997mm,Polyakov:2003dx,Diakonov:2003jj} is an
attractive possibility. The bump structure in the $n\eta$ invariant
mass spectrum was confirmed by the CB-ELSA/TAPS \cite{Jaegle:2005fa}
and LEPS \cite{Hosaka:2007} collaborations.

Different interpretations have been offered as origin of this
structure. Choi {\it et al.} \cite{Choi:2005ki} use three known
nucleon resonances,  $N(1535)S_{11}$, $N(1650)S_{11}$
$N(1710)P_{11}$, and a narrow state at 1675 MeV which they discuss
as pentaquark $N(1675)P_{11}$. Vector meson exchange in the $t$
channel was used as a background amplitude.

The Giessen group arrives at different conclusions
\cite{Shklyar:2006xw}. Within their unitary coupled-channels
effective Lagrangian approach,  the cross section of $\eta$
photoproduction on the neutron was fully described. The  peak at
$\sqrt{s}$=1.66 GeV was explained as coupled-channel effect due to
$N(1650)S_{11}$ and $N(1710)P_{11}$ resonance excitations. No narrow
resonance was required.

The analysis of Fix {\it et al.} \cite{Fix:2007st} required, in
addition to the conventional ingredients of the MAID model, a narrow
state which was assumed to have $P_{11}$ quantum numbers.

In this paper, we present a partial wave analysis of the recent data
of the CB-ELSA/TAPS collaboration \cite{Jaegle:2008ux} on $\gamma
d\to p_{\rm spectator} n\eta$. In view of the long-standing
discrepancies between the photo-production amplitude $A^n_{1/2}$ for
$N(1535)S_{11}$ production ($A^n_{1/2}=-0.020\pm
0.035$\,GeV$^{-1/2}$ from $\gamma n\to n\pi^0$ \cite{Arndt:1995ak};
$A^n_{1/2}=-0.100\pm 0.030$\,GeV$^{-1/2}$ from $\gamma n\to n\eta$
\cite{Krusche:1995zx}), it seems adequate to include some older data
on $\gamma n\to n\pi^0$, but we also include recent data from GRAAL
on the beam asymmetry for $\gamma n\to n\eta$ \cite{Fantini:2008zz}
and $\gamma n\to n\pi^0$ \cite{LeviSandri:2007ku}. In addition, we
use data on photoproduction of $2\pi^0$, $\pi^0\eta$ and of hyperons
as well as some partial-wave amplitudes from elastic $\pi N$
scattering and data on $\pi^-p\to p2\pi^0$. A survey of the data
used in the fits, of the partial wave analysis method and of recent
results can be found elsewhere
\cite{Thoma:2007bm,:2007bk,Anisovich:2007bq,Nikonov:2007br,Horn:2008}.

The paper is organized as follows: after this introduction we
present, in section \ref{Fermi}, how the Fermi motion of the neutron
in the deuteron is treated.  Reasonable consistency is found for the
$\gamma p\to p\eta$ cross sections for protons bound in deuterons -
folded with the Fermi momentum distribution - with the cross
sections measured on free protons. The success encouraged us to
perform a partial wave analysis for the part of the data where the
proton acts as a spectator. The fits and the results are presented
in section 3. The paper ends with a short summary and our
conclusions (section 4).

\section{\label{Fermi}Fermi motion}
Experimentally, the cross section for $\eta$ meson photoproduction
off  deuterons is measured. The deuteron is at rest in the
laboratory system, the neutron not. It has had, at the moment of the
interaction, the same (but opposite) momentum as the proton. A cut
in the missing momentum of the (undetected) proton selects events in
which the proton acted as spectator.

There are two approaches to fit data. First, one could unfold the
experimentally observed cross section to determine the cross section
for a neutron target. This data can then be fitted. Alternatively,
the calculated cross section can be folded with the neutron
momentum. In this way, adopted here, the fitted cross section can be
compared directly to the measured quantities.

The differential cross section for production of $n$ particles in
the photon nucleon interaction has the form
\be d\sigma_{\gamma N}=\frac{(2\pi)^4
|A|^2}{4\,\sqrt{(k_1 k_2)^2-m_1^2m_2^2}}\,
d\Phi_n(P,q_1,\ldots,q_n)
\label{xsec_elem}
\ee
where $k_i$ and $m_i$ are the four--momenta and masses of the
initial particles, $P$ is the total momentum ($P=k_1+k_2$) and $q_i$
are the four--momenta of final state particles. The
$d\Phi_n(P,q_1,\ldots,q_n)$ is the n-body phase volume
\be
d\Phi_n(P,q_1,\ldots,q_n) = \delta^4(P- \sum\limits_{i=1}^n q_i)
\prod\limits_{i=1}^n \frac{d^3q_i}{(2\pi)^3 2q_{0i}}
\label{phv}
\ee
where $q_{0i}$ are the energy components.

In the case of meson photoproduction off nucleons bound in a
deuteron, the cross section (\ref{xsec_elem}) should be integrated
over its momentum:
\be
d\sigma_{\gamma D} =\!\int\! d|\vec p_N| \;|\vec p_N|^2 f^2(|\vec p_N|)
\frac{dz_N d\phi_N}{4\pi}
\frac{(2\pi)^4
|A|^2}{4\,\sqrt{(k_1 p_N)^2}}\times
\nn \\
d\Phi_n(k_1+p_N,q_1,\ldots,q_n)\,,~~~ \label{xsec_int} \ee where
$\vec p_N$ is the momentum of the nucleon, $z_N=\cos \Theta_N$, and
where $dz_N d\phi_N$ forms the solid angle element of the nucleon in
the laboratory system. The function $f(P_N)$ describes the momentum
distribution of the nucleon inside the deuteron. It can be chosen in
the form of the Paris \cite{Lacombe:1981eg} or Gatchina wave
function \cite{as_wf}.

The spectator nucleon in the $\gamma d$ interaction has the momentum
$p_{s}=-p_{N}$ in the lab system (deuteron at rest) and is on shell.
Therefore the energy of the interacting particle is given by
 \be
E_N=M_d-\sqrt{m_s^2+\vec p^2_N}
 \ee
where $M_d$ is the deuteron mass and $m_s$ is the mass of the
spectator nucleon. The off shell mass squared ($t$) of the
interacting nucleon and the total energy squared in the $\gamma N$
interaction $s_{tot}(t)$ are equal to
 \be
 t=E_N^2-m_N^2\qquad s_{tot}(t)=t+2E_\gamma\big(\sqrt{\vec p_N^2+t}-|\vec p_N|z_N\big
). \nonumber
 \ee

A major problem in this approach is to relate the off-shell
amplitude of the interacting particles with measurable on-shell
distributions. It can be shown \cite{Krusche:1995zx} that the best
description is achieved under the assumption
 \be
\sigma(s_{tot}(t),t)&=&\sigma(s_{tot}(m_N^2),m_N^2)\nn\\
\sigma(s_{tot}(t),t)&=&0 \quad\mbox{for}\quad s_{tot}(t)<(m_N+m_\eta)^2\,.
\label{on_shell}
 \ee
Due to relation (\ref{on_shell}), all further calculations can be
performed for an on-shell nucleon. The components of the initial
4-vectors in the lab system are defined as
\be
&&p_N = (p_{0N}, p_{xN}, p_{yN}, p_{zN})\quad k_1 = (E_\gamma, 0, 0,
E_\gamma)
\nn \\
&&p_{xN}=|\vec p_N|\sin \Theta_N \cos \phi_N,\quad
p_{zN}=|\vec p_N|z_N\quad
\nn \\
&&p_{yN}=|\vec p_N|\sin \Theta_N
\sin \phi_N, \quad  p_{0N} =\sqrt{m_N^2+ \vec p_N^2}
\ee
where $m_N$ is nucleon mass and momentum of the photon $\vec k_1$ is
directed along $z$-axis.

In the case of single-meson photoproduction, the amplitude depends
on the total energy of the $\gamma N$ system and the angle between
the initial photon and the final meson calculated in the center of
mass system (cms) of the reaction. Differential cross sections are
usually given in the center of mass system of the photon and the hit
nucleon. We will call this system as ``data'' system. Let us
calculate the momentum of the particles and scattering angles in the
laboratory system, in the cms and the ``data'' system.

The total energy squared (which is an invariant value) can be
calculated, for example, in the laboratory system:
\be
s_{tot} = (k_1 + p_N)^2=m_N^2 + 2E_\gamma (p_{0N}- |\vec p_N| z_N ).
\ee

Then in cms of the reaction:
\be
z_{cms} = \frac{\vec q^{cms}_1 \vec k^{cms}_1}{|\vec q^{cms}_1|
|\vec k^{cms}_1|} = \frac{q^{cms}_{10} k^{cms}_{10} - (q_1
k_1)}{|\vec q^{cms}_1| |\vec
k^{cms}_1|} \;, \nn \\
q_{10}^{cms} = \frac{s_{tot} +m_1^2 -m_N^2}{2\sqrt{s_{tot}}}\;,\;\;
k_{10}^{cms} = \frac{s_{tot} -m_N^2}{2\sqrt{s_{tot}}}\;, \nn\\
|\vec q^{cms}_1| = \sqrt{(q_{10}^{cms})^2 - m_1^2}\;,\;\;
\qquad|\vec k^{cms}_1| = k_{10}^{cms}\,.
\label{z_cms}
\ee
Here, $q_1$ is the 4-momentum of the final meson with mass $m_1$.
The invariant quantity $(q_1 k_1)$ can be calculated in any system
(e.g. the ``data'' system).

To define the photon 4-vector in the ``data'' system let us
calculate the invariant $(k_1 P_{\rm eff})$, where $P_{\rm eff}$ is
the sum of photon and nucleon momenta, in both the laboratory system
and in the ``data" system.
 \be
 &P_{eff}^{lab} = (m_N+E_\gamma,0,0,E_\gamma),\;\;
 &k_1^{lab} = (E_\gamma, 0,0,E_\gamma) \nn \\
 &P_{eff}^{data} = (\sqrt{s_{eff}},0,0,0),\;\; &k_1^{data} =
(E^m_\gamma, 0,0,E^m_\gamma)\,.~~~~~
 \ee
Comparing this invariant in the two systems we obtain
 \be
 E^m_\gamma
 = \frac{m_N E_\gamma}{\sqrt{s_{eff}}}, \qquad s_{eff} = m_N^2 + 2m_N
 E_\gamma
 \ee
Then the invariant $(k_1 q_1)$ in the ``data'' system is equal to
 \be
 q_1 k_1 = \frac{m_N E_\gamma}{\sqrt{s_{eff}}} (\sqrt{m_1^2+|\vec
q|^2} - |\vec q|\,z).
 \ee
In this equation, $\vec q$ is the laboratory momentum of the meson.
As a result, we can express all variables in terms of measured
values; needed are - in the ``data'' system - the photon energy
$E_\gamma$ and $z$, the cosine of the angle between meson and
photon.

Our next task is to define the initial nucleon momentum in the
``data'' system. Then:
 \be
&p_N^m& = (p_{0N}^m, p_{xN},p_{yN},p_{zN}^m)\nn \\
&p_{0N}^m &= \frac{p_{0N} m_N + E_\gamma(p_{0N}-p_{zN})}{\sqrt{s_{eff}}}\,,
\nn\\
&p_{zN}^m& = p_{0N}^m - (p_{0N} -p_{zN})\frac{\sqrt{s_{eff}}}{m_N}
 \ee
The transition from the lab system to the ``data'' system is
performed via a boost along the $z$-axis, so "$x$" and "$y$"
components of 4-vector are not changed. The equations for $p_{0N}^m$
and $p_{zN}^m$ can be obtained from the invariance of the scalar
products $(p_N P_{eff})$ and $(p_N k_1)$ calculated in the lab and
``data'' systems.

Now one can calculate the phase volume for the meson-nucleon final
state in the ``data'' system:
 \be
d\Phi_n(k^m_1\!+\!p_N^m,q_1,q_2) =
\frac{\delta(P^m_0\!-q_{01}\!-q_{02} )\, d^3q_1}{4(2\pi)^6
q_{10}q_{20}}\,
\label{phv_m0}
 \ee
where
 \be
&&q_{01}=\sqrt{m_1^2+|\vec q|^2}\qquad
q_{02}=\sqrt{m_N^2+|\vec P^m-\vec q|^2}\nn \\
&&P^m=(p_{0N}^m+E^m_\gamma, p_{xN},p_{yN},p_{zN}^m+E^m_\gamma)
 \ee
and $m_1$ is the mass of the final meson.

From energy conservation, the  absolute value of the meson momentum
in the ``data'' system is calculated to
 \be
 |\vec q| =  \frac{\Sigma \xi|\vec P^m| + P^m_0 \sqrt{\Sigma^2 -
m_1^2\left [(P^m_0)^2 -|\vec P^m|^2\xi\right ]}} {(P^m_0)^2 - |\vec
P^m|^2 \xi^2}\,~~~ \label{q_abs}
 \ee
where
 \be
\Sigma = \frac12 (s_{tot} +m_1^2 -m_N^2)
 \ee
and $\xi$ is the cosine of the angle between $\vec P^m$ and $\vec
q_1$:
 \be
\xi = \frac
{zP^m_z + |\vec p_N|\sqrt{1-z_N^2}\sqrt{1-z^2} cos(\phi_N -
\phi)}{|\vec P^m|}
\label{xi_phi}
 \ee
and where $\phi$ is the azimuthal angle of the final meson. Then the
phase volume is given by
 \be
 d\Phi_n(P^m,q_1,q_2)\!=\!\frac{1}{4(2\pi)^6} \frac{|\vec
q|^2dzd\phi} {|\vec q| P^m_0 - |\vec P^m| \xi \sqrt{m_1^2\!+\!|\vec
q|^2}} \label{phv_m}.~~~ \ee

All variables now depend only on the relative angle $\phi_N-\phi$.
In the evaluation of the cross section (\ref{xsec_int}), one
integration can be performed trivially and eq. (\ref{xsec_int}) can
be rewritten in the form
 \be
 d\sigma_{\gamma D} = \int d|\vec p_N| \,|\vec p_N|^2 f^2(|\vec
p_N|) \frac{dz_N d\phi_N}{4\pi}\times
\nn \\
\frac{|A(s_{tot}, z_{cms})|^2}
{32\pi\,(k_1 p_N)}
\frac{|\vec q|^2 dz}{|\vec q| P^m_0 - |\vec P^m| \xi \sqrt{m_1^2 + |\vec q|^2}}
\label{cross_sec2}
 \ee
where
 \be
\xi &=&\frac{zP^m_z + |\vec p_N|\sqrt{1-z_N^2}\sqrt{1-z^2} cos(\phi_N)}
{|\vec P^m|}
\nn \\
P^m_0 &=& \frac{(p_{0N} +E_\gamma) m_N + E_\gamma(p_{0N}-p_{Nz})}{\sqrt{s_{eff}}}
\nn \\
|\vec P^m| &=& \frac{m_N(p_{zN} +E_\gamma) - E_\gamma
(p_{0N}-p_{zN})}{\sqrt{s_{eff}}} + E_\gamma~~
\label{xi_fin}
 \ee
and $|\vec q|$ is defined by eq.(\ref{q_abs}).

At low energies, the phase volume given by eq.(\ref{phv_m})
decreases for forward angles leading to a corresponding behavior of
the cross section in the region where the $S_{11}$ wave is the
dominant contribution.

\section{Photoproduction of $\eta$-mesons off deuterons}

\subsection{$\eta$ photoproduction off protons}

\begin{figure}[ph]
\centerline{
\epsfig{file=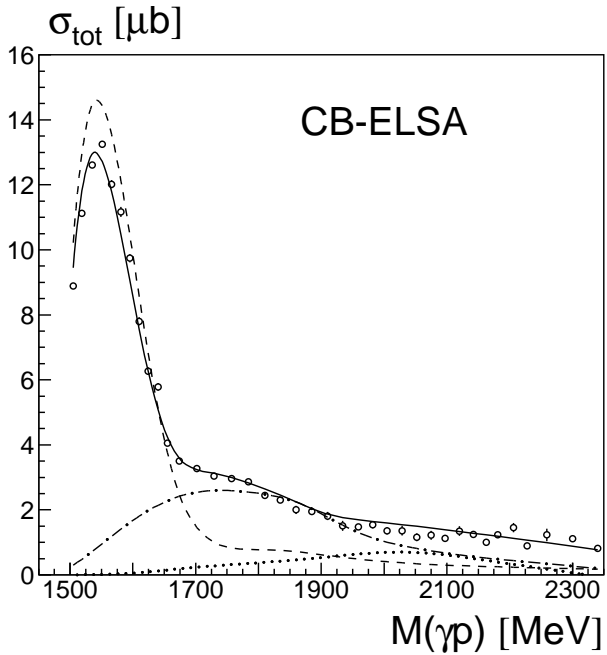,width=0.45\textwidth,clip=}}
\caption{The total cross section for $\rm\gamma p \rightarrow \eta
p$ from the deuteron target. The description of the data (solid
line) is obtained from the solution on the free proton smeared with
the Paris wave function. The dashed line is the $S_{11}$, the
dash-dotted line the $P_{13}$, and the dotted line the $D_{15}$
contribution.} \label{pr_tot_1} \vspace{-4mm}
\end{figure}

\begin{figure}[pt]
\centerline{
\epsfig{file=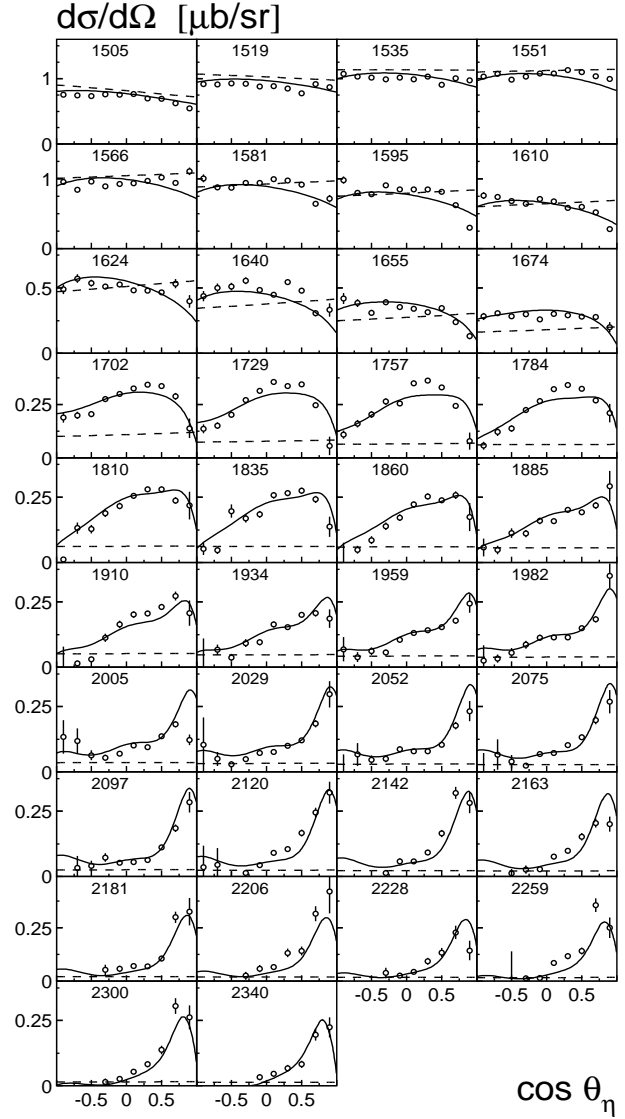,width=0.45\textwidth,height=0.655\textheight,clip=}
\vspace{-4mm}} \caption{The differential cross section for
$\rm\gamma p \rightarrow \eta p$ from the deuteron target in the
1505-2340\,MeV mass range. The solid curves represent a fit to
free-proton data smeared with the Paris wave function. The dashed
curves show the contribution of the $S_{11}$ wave.\vspace{-2mm}}
\label{pr_dcs_1}
\end{figure}

As a first step, we describe $\eta$ photoproduction off protons
bound in a deuteron. This allows us to test the reliability of the
folding procedure accounting for the Fermi motion. The total cross
section and the angular distributions are presented in
Figs.\ref{pr_tot_1} and \ref{pr_dcs_1} for the case of the Paris
wave function. The error bars on these figures represent statistical
errors only.

For the fits, we use alternatively the Paris wave function
\cite{Lacombe:1981eg} or a deuteron wave function obtained from a
dispersion $N/D$-method \cite{as_wf}. The fit uses no new
parameters: all masses, widths, partial decay widths, and helicity
amplitudes are determined by the data from $\gamma p\to p\eta$
\cite{Bartholomy:2007zz} and the fits described in
\cite{Anisovich:2007bq}. The $\chi^2$ was found to be $2396$ for 380
points using the Paris wave function and $2410$ with the $N/D$-based
wave function. The $\chi^2$'s are similar demonstrating that the
extraction of cross section from deuteron data is insensitive to
details of the deuteron wave function. This observation is confirmed
when cross sections for $\gamma n\to n\eta$ are extracted. Hence we
show here only figures obtained by using the Paris wave function.
The $\chi^2$'s are large; inspecting the differential cross sections
and the deviations between data and fit suggests that systematic
errors in the extraction of the cross sections may be responsible
for a significant fraction of the large $\chi^2$'s.\vspace{-3mm}

\subsection{The $\eta$ and $\pi$ photoproduction off neutrons}

In the present analysis, the following data sets are added to our
data base used in our fits: $\eta$ photoproduction off the neutron
from the CB-ELSA experiment \cite{Jaegle:2008ux}, beam asymmetry for
$\eta$ \cite{Fantini:2008zz} and $\pi$  \cite{LeviSandri:2007ku}
photoproduction off the neutron from the GRAAL experiment and $\pi$
photoproduction off the neutron from the SAID database \cite{SAID}.
These data were fitted together with other photo- and pion-induced
single and double photoproduction data as listed in the
Introduction. All our fits produced a very similar $\chi^2$ for the
Paris and $N/D$-based wave function. Hence we discuss only the
investigations which had been done using the Paris wave functions.

\begin{figure}[pt]
\centerline{
\epsfig{file=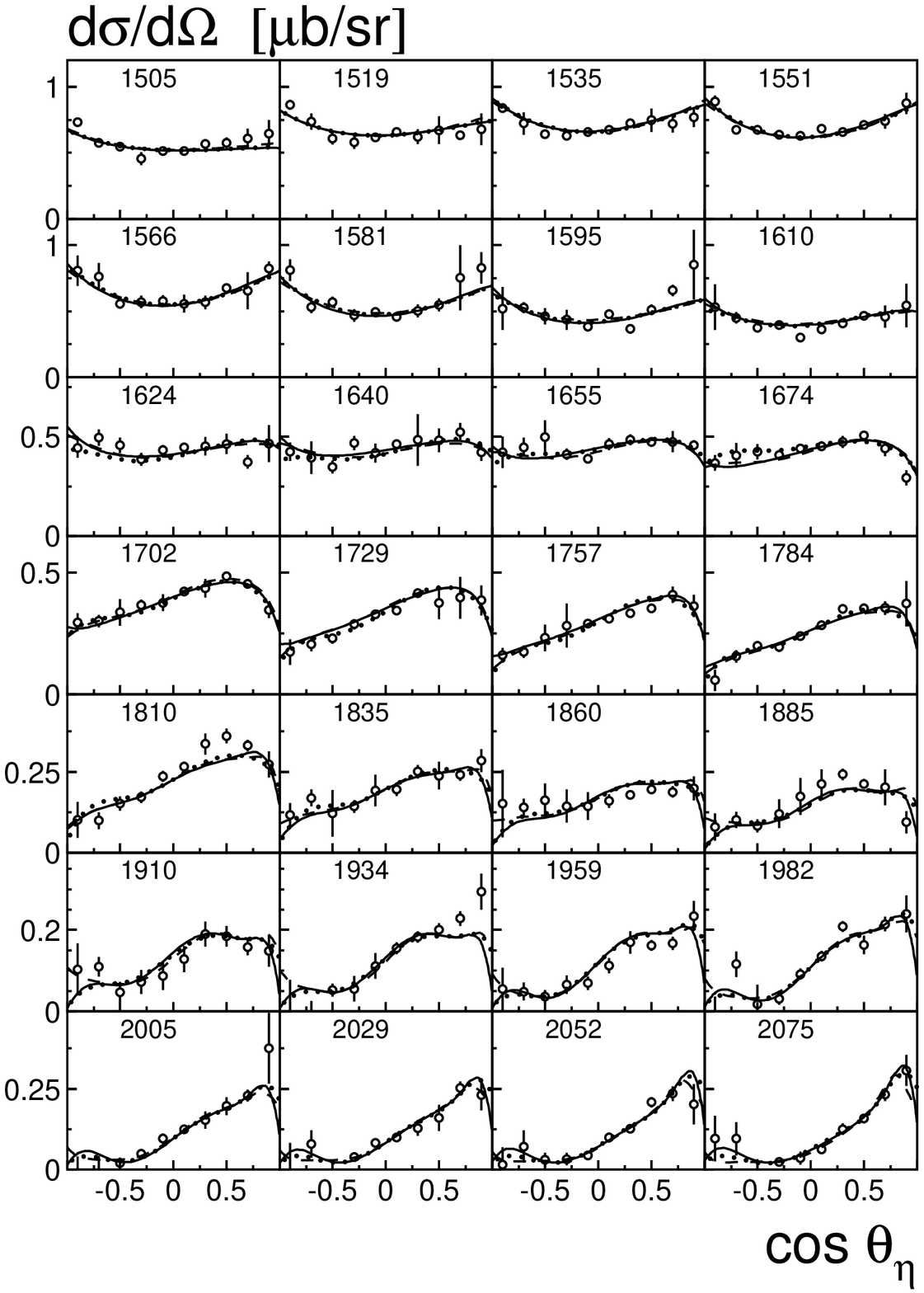,width=0.45\textwidth,clip=}}
\caption{\label{etap_dif}The differential cross section for
$\rm\gamma n \rightarrow \eta n$ off deuterons \cite{Jaegle:2008ux}.
The PWA description is shown as solid line (solution 1), dashed line
(solution 2) and dotted line (solution 3). \vspace{3mm}}
%\end{figure}
%\begin{figure}[pt]
\centerline{
\epsfig{file=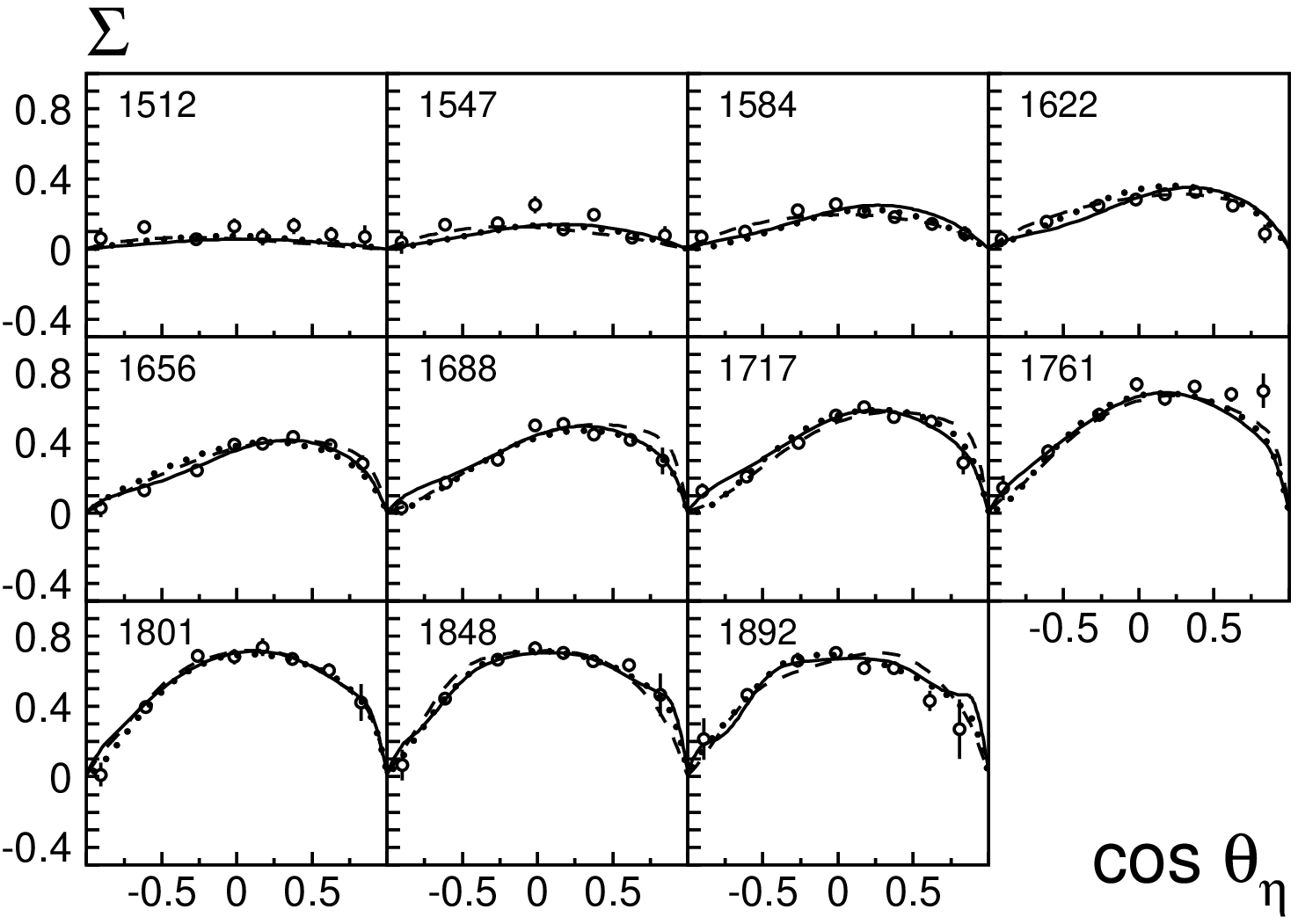,width=0.48\textwidth,clip=}}
\caption{\label{etap-Sigma}Beam asymmetry for $\rm\gamma n
\rightarrow \eta n$ for neutrons bound in a deuteron
\cite{Fantini:2008zz}. The PWA description is shown as solid line
(solutions 1), dashed line (solution 2), and dotted line (solution
3). \vspace{3mm}}\normalsize The differential cross section for
$\gamma n\to n\eta$ is shown in Fig.~\ref{etap_dif}, the beam
asymmetry in Fig.~\ref{etap-Sigma}. The corresponding data for
$\gamma n \rightarrow \pi^0 n$ are shown in Fig.~\ref{pi0_dif} and
\ref{pi0_Sigma}. The data are fitted using three different
scenarios.  In all cases, the most significant contributions came
from the $S_{11}$, $P_{11}$, and $P_{13}$ partial waves, with
$S_{11}$ providing the largest contribution.\end{figure}

\begin{figure}[pt]
\centerline{
\epsfig{file=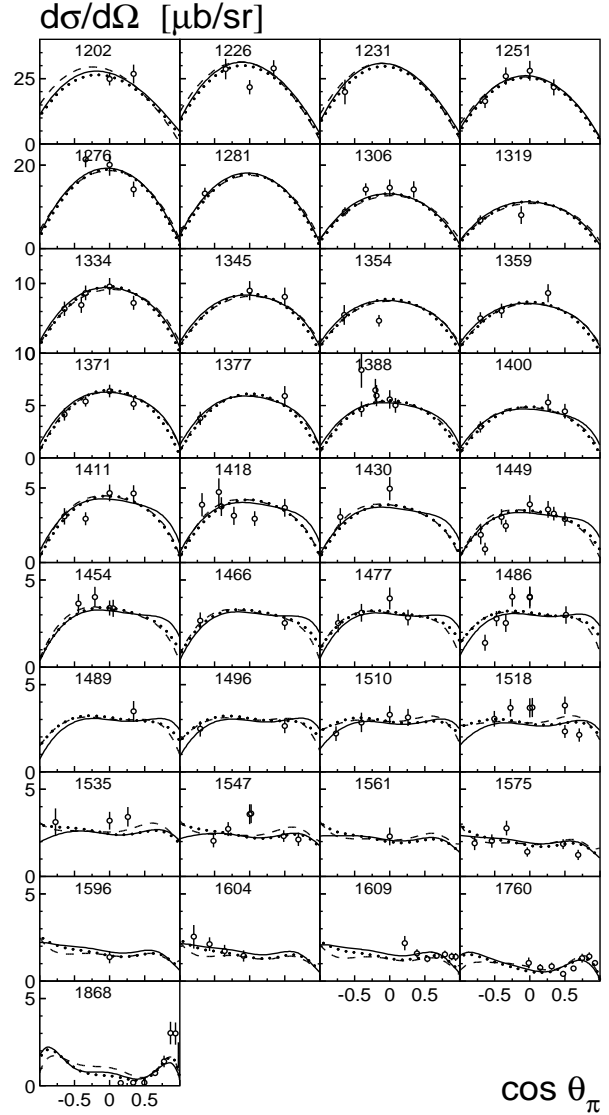,width=0.45\textwidth,height=0.655\textheight,clip=}}
\caption{\label{pi0_dif}The differential cross section for
$\rm\gamma n \rightarrow \pi^0 n$ using a deuteron target
\cite{SAID}. The PWA description is shown as solid line (solutions
1) , dashed line (solution 2), and dotted line (solution 3).}
\end{figure}\noindent
These three partial waves, and for the waves $P_{33}$ and $D_{33}$
which are irrelevant here, were described using K-matrices. For the
other less important waves, relativistic multi-channel Breit-Wigner
amplitudes were used. For the important waves, the elastic
scattering amplitudes from \cite{Arndt:2006bf} were included in the
fit using the same K-matrix as for the photoproduction data.

\begin{figure}[pt]
\centerline{
\epsfig{file=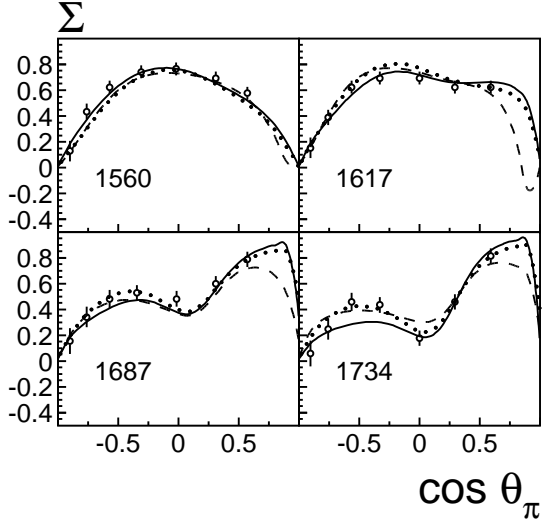,width=0.40\textwidth,clip=}\vspace{-2mm}
} \caption{\label{pi0_Sigma}Beam asymmetry for the reaction
$\rm\gamma n \rightarrow \pi^0 n$ from the deuteron target
\cite{LeviSandri:2007ku}. The PWA description is shown as solid line
(solutions 1), dashed line (solution 2), and dotted line (solution
3).\vspace{-3mm}}
\end{figure}

\begin{table}[pb] \vspace{-3mm}\caption{\label{list_chi} Single meson
photoproduction off neutron data used in the partial wave analysis
and $\chi^2$ for solutions 1 (interference in $S_{11}$ wave), 2
($N(1710)P_{11}$), and 3 (narrow $P_{11}$).}
\renewcommand{\arraystretch}{1.2}
\begin{center}
\begin{tabular}{lccccr}
\hline\hline
&&&&&\vspace*{-3mm}\\
Observable\hspace*{-3mm}&\hspace*{-3mm}$N_{\rm data}$\hspace*{-3mm}&
\hspace*{-3mm}$\frac{\chi^2}{N_{\rm
data}}$\hspace*{-3mm}&\hspace*{-3mm}$\frac{\chi^2}{N_{\rm data}}$
\hspace*{-4mm}&\hspace*{-4mm}$\frac{\chi^2}{N_{\rm
data}}\hspace*{-3mm}$&
 Ref.\vspace*{2mm} \\
\hline \hline
& & Sol. 1 &Sol. 2 &  Sol. 3& \\
$\rm\sigma(\gamma n \rightarrow n\eta)$\hspace*{-3mm}
 &  280 &  1.32  &  1.26   & 1.31&
 \cite{Jaegle:2008ux}\\
$\rm\Sigma(\gamma n \rightarrow n\eta)$\hspace*{-3mm}
 &  88 &   1.75  &  1.85   & 1.79&
 \cite{Fantini:2008zz}\\
$\rm\sigma(\gamma n \rightarrow n\pi^0)$\hspace*{-3mm}
 &  147 &   2.01  &  2.35   & 2.03&
 \cite{SAID}\\
$\rm\Sigma(\gamma n \rightarrow n\pi^0)$\hspace*{-3mm}
 &  28 &   1.02  &  1.07   & 0.90&
 \cite{LeviSandri:2007ku}\\
 \hline \hline
\end{tabular}
\end{center}
\renewcommand{\arraystretch}{1.0}
\vspace{-0.3cm}
\end{table}

In the first solution, the low-energy region is described mainly by
the interference between $N(1535)S_{11}$ \hfill and $N(1650)S_{11}$.
In the second solution we enforce a large contribution from a
standard $N(1710)P_{11}$ resonance. In the third solution, we test
the possibility of a narrow (less than 10 MeV) state at about
1650\,MeV. The resulting fit curves are also shown in
Figs.~\ref{etap_dif}-\ref{pi0_Sigma}. In Table \ref{list_chi} we
give a breakdown of the $\chi^2$ contributions of the four data sets
in the three scenarios. All three scenarios provide an adequate
description of the dip-bump structure observed in the $\gamma n\to
n\eta$ total cross section.

A first analysis \cite{Anisovich:2007} of the preliminary CB-ELSA
data \cite{Jaegle:2005fa} presented at NSTAR 2007 did not include
$t$ and $u$-exchanges due to the fact that in the low energy region
these contributions are difficult to separate from other
non-resonant terms. The present analysis is extended up to 2.1 GeV,
first without contributions from $t$ and $u$ channel exchanges and
second with these contributions included. The fits with $t$ and $u$
exchanges result in a slightly better description of the high energy
tail but qualitatively do not change the solutions in the region
below 1.75\,GeV. However, both, the inclusion of $t$ and $u$ channel
exchanges and the use of the final data decreased the helicity
amplitudes. The new values reported here supersede those reported at
NSTAR 2007. The contributions of high mass states are ambiguous and
cannot be identified reliably. More data and further systematical
investigations are needed. These uncertainties do not affect our
conclusions concerning the low-mass region which is the prime issue
of the study.

\subsubsection{Parameterization of the $S_{11}$ wave}

Following our previous analyses
\cite{Bartholomy:2007zz},\cite{Thoma:2007bm} the $S_{11}$ wave was
parameterized as two pole, 5 channel K-matrix amplitude:
\be
 K_{ab}\;=\;\sum\limits_{\alpha=1}^{2} \frac{g_a^{(\alpha)} g_b^{(\alpha)}}
{M^2_\alpha - s} \;+\; f_{ab},
\ee
where $a,b=$$p \pi$, $p\eta$, $K\Lambda$, $K\Sigma$, $\Delta\pi$,
and $M_\alpha$ and $g_a^{(\alpha)}$ are masses and coupling
constants of the K-matrix poles.

In \cite{Arndt:1995bj} the non-resonant contributions were
parameterized as linear mass dependent functions. We also found that
such mass dependence introduced for the $\pi N\to \pi N$ and $\pi
N\to \eta N$ and $\eta N\to \eta N$ non-resonant terms improves
notably the description of the pion induced and photoproduction
reactions. However, in our parameterization we introduced in
addition a factor which suppresses the divergency of the
non-resonant terms at large energies. Thus
\be
f_{ab}=(f^{(1)}_{ab}+f^{(2)}_{ab}\sqrt{s})\frac{2+s_{ab}}{s+s_{ab}}
\qquad a,b=\pi N,\eta N
\ee
and $s_{ab}>0$. The non-resonant transitions between $\pi N\to
K\Lambda$, $\pi N\to K\Sigma$ and $\pi N\to \Delta \pi$ channels
also improve the combined description. However these terms can be
parameterized as constants. All other transitions contribute very
little to the data description and were fixed to zero.

The amplitude for the transition between K-matrix channels can be
written as: \be A_{ab} \;=\; \hat K_{ac}\;(\hat I\;-\;i\hat \rho
\hat K)^{-1}_{cb}\,. \ee The phase space $\hat \rho$ is a diagonal
matrix $\rho_{ab}\;=\;\delta_{ab}\;\rho_a$ with
 \be
\rho_{a}(s)=\frac{2|\vec k_B|}{\sqrt{s}}\,
\frac{m^a_B+\sqrt{(m^a_B)^2+|\vec k_B|^2}}{2m_B^a} \ee for the two
body final states. Here $m^a_B$ is the mass and $\vec k_B$ is the
momentum (calculated in the c.m.s. of the reaction) of the baryon in
the channel (a) (see \cite{Anisovich:2006bc}). The parameterization
of the $\Delta \pi$ phase volume is given in details in
\cite{Anisovich:2006bc}.

\begin{figure}[pt]
\centerline{ \epsfig{file=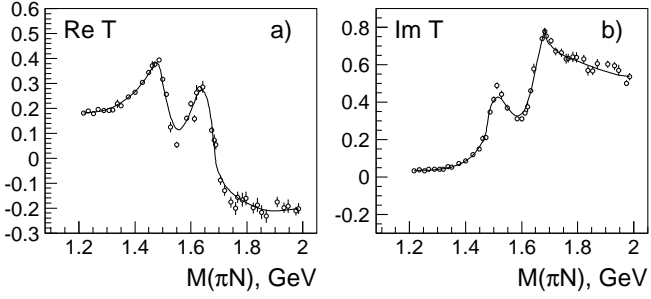,width=0.48\textwidth,clip=}}
\caption{The description of the $\pi N\to \pi N$ $S_{11}$ amplitude
obtained in the combined solution. The data are taken from energy
independent solution \cite{Arndt:2006bf}.}
\label{s11_pin}
\end{figure}

\begin{figure}[pt]
\centerline{ \epsfig{file=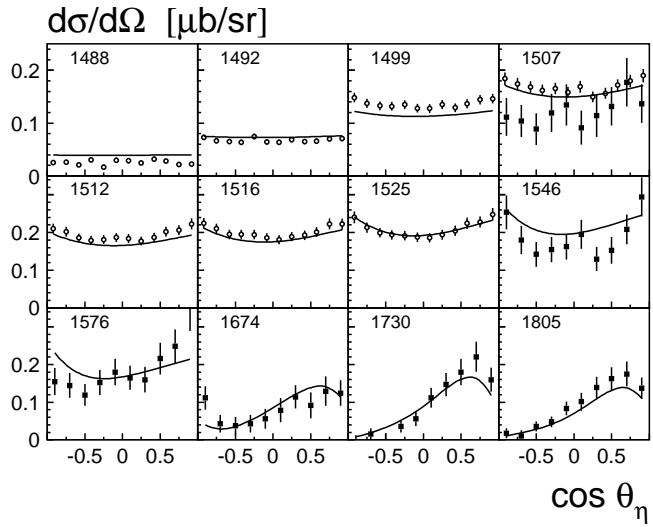,width=0.48\textwidth,clip=}}
\caption{The description of the $\pi p\to \eta n$ differential cross
section obtained in the combined solution. The data are taken from
\cite{Prakhov:2005qb} (open circles) and \cite{Richards:1970cy}
(full squares).}
\label{s11_pi_eta}
\end{figure}

The K-matrix parameters for the $\pi N$ and  $\eta N$ channels are
constrained from the fit of the elastic $\pi N\to \pi N$ data
(extracted by \cite{Arndt:2006bf}) and the fit of the $\pi^- p\to
\eta n$ differential cross section
\cite{Prakhov:2005qb},\cite{Richards:1970cy}. The description of
these data is shown in Figs.~\ref{s11_pin} and \ref{s11_pi_eta}.

The photoproduction amplitude is parameterized in the $P$--vector
approach since the $\gamma\rm N$ couplings are weak and do not
contribute to rescattering. The amplitude is then given by
\be
A_a \;=\; \hat P_b\;(\hat I\;-\;i\hat \rho \hat K)^{-1}_{ba}\,.
\ee
with $P$-vector parameterized as:
\be
P_{b}\;=\;\sum_\alpha \frac{ g_{\gamma \rm N}^{(\alpha)}
g_b^{(\alpha)}}{M^2_\alpha - s} \;+\; \tilde f_{b}
\ee
Here $g_{\gamma\rm N}^{(\alpha)}$ are $\gamma N$ couplings of the
K-matrix poles and $\tilde f_{b}$ are non--resonant production
terms, parameterized in the fit as real constants.

\subsubsection{Interference in the $S_{11}$ wave}
\begin{figure}[pt]
\centerline{
\epsfig{file=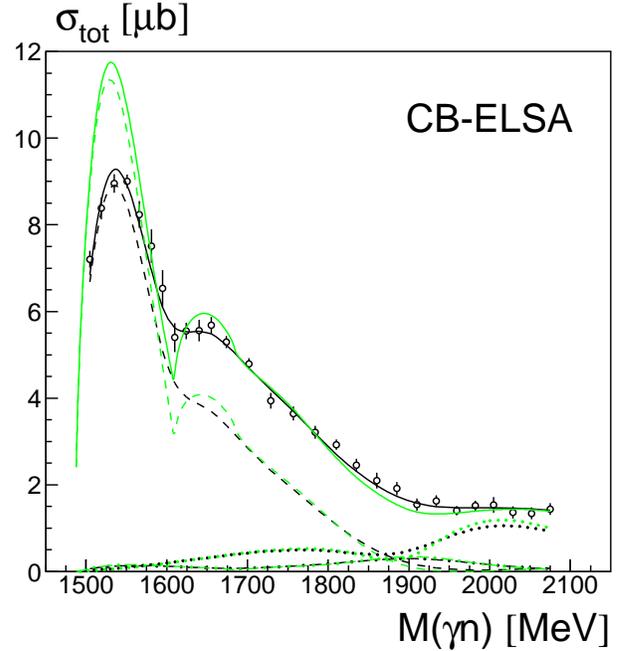,width=0.45\textwidth,clip=}}
\caption{The total cross section for the reaction $\rm\gamma n
\rightarrow \eta n$ from the deuteron target \cite{Jaegle:2008ux}.
The PWA description from Solution 1 (Paris wave function) is shown
as the solid line. The dashed line is the $S_{11}$ contribution,
dash-dotted line is $P_{11}$ contribution and dotted line is
$P_{13}$ contribution The grey (online: green) curves show the
corresponding cross sections on the free neutron (no Fermi motion)}
\label{eta_s11_tot}
\end{figure}

The first solution with a strong interference in the $S_{11}$ wave
provides a very good description of the fitted data (see Table
\ref{list_chi}). In particular the bump in the 1650 MeV region is
well described. This solution gives the following helicity couplings
for $S_{11}$ resonances calculated at the pole positions of the
$S_{11}$ amplitude:
 \be
 S_{11}(1535):\; & A^n_{1/2}\! =\! -0.080 \pm 0.020\;,\;& \phi \!=\! 12^\circ \pm
 10^\circ \nonumber\\
 S_{11}(1650):\; & A^n_{1/2}\! = \! -0.060\pm 0.015\;,\;& \phi \!=\! 40^\circ \pm
 25^\circ ~~~~~~
 \label{hel_res}
 \ee

The bump in the region of 1650\,MeV appears due to an interference
between $S_{11}(1535)$, $S_{11}(1650)$  and a non-resonant
background. In our combined solution of the single photoproduction
data $S_{11}(1650)$ has  the rather small (~15\%) branching ratio
into the $\eta N$ channel. Therefore an appreciable large coupling
of this state to the $\gamma n$ channel is needed to describe the
bump structure. Here we are in contradiction with the Giessen result
\cite{Shklyar:2006xw} where the bump appears with decreasing of the
$S_{11}(1650)$ $\gamma n$ helicity coupling.  The two $S_{11}$
states have very close (apart from overall sign) couplings into
$\gamma p$ and $\gamma n$ channels. However there is an important
correlation: the phase difference between the couplings is fixed
more precise than the absolute numbers. We found the phase
difference $5\pm 5$ degrees for the $\gamma p$ channel and $28\pm 8$
for the $\gamma n$ channel. The bump structure in the 1650-1700\,MeV
region becomes much less pronounced in the case of a smaller phase
difference (see solutions discussed below).

The K-matrix parameters of the $S_{11}$ wave are rather firmly fixed
from the fit of the elastic data and photoproduction reactions off
the proton. The only mandatory parameters to fit $\gamma n$
reactions are two P-vector $\gamma n$ pole couplings and five
non-resonant production constants. The $\gamma n\to \pi n$ and
$\gamma n\to \eta n$ can be fixed directly from the combined
analysis of the differential cross sections and beam asymmetry data
from the neutron target. Fixing these parameters to zero leads to a
large deterioration of the combined description.

Among other non-resonant contributions the most important one is the
direct production of the $K\Lambda$ channel. It can notably
influence the structure at 1650-1700 MeV which is situated in
vicinity of the $K\Lambda$ threshold. The $K\Sigma$ production only
slightly improves the description at high energies and $\Delta\pi$
can be put to zero.

To check the influence of the $\gamma n\to K\Lambda$ and $\gamma
n\to K\Sigma$ direct production terms we performed the fit fixing
these parameters to zero value. To reproduce the description of the
$\gamma n\to \eta n$ data we increase the weight of this data set by
a factor of 2. In this fit we could reproduce the unpolarized $\eta
n$ cross section and beam asymmetries for $\pi^0 n$ and $\eta n$.
However the fit failed to reproduce the unpolarized $\gamma n\to
\pi^0n$ differential cross section: the $\chi^2$ changed from 2.11
to 2.71. The residue for the $S_{11}(1535)$ state did not change
within errors (\ref{hel_res}). The helicity coupling of the
$S_{11}(1650)$ is slightly bigger in this solution: $\sim 0.090$
GeV$^{-\frac 12}$ and the phase difference with the first pole
coupling reached $120$ degrees.

A simplified parameterization provides a simplified picture: the
difference in phases of helicity couplings in the $\gamma p$ and
$\gamma n$ reactions is clearly seen. However it failed to describe
simultaneously all reactions. This is one of the main reasons why
analyses of different sets of photoproduction data results in
incompatible helicity couplings.

The $P_{11}$ and $P_{13}$ waves provide contributions of similar
strengths to $n\eta$. In the 1700\,MeV region, the $N(1710)P_{11}$
resonance is weak while $N(1720)P_{13}$ makes a small contribution.
The $P_{11}$ wave becomes stronger at 1900\,MeV.

\subsubsection {Enforcing $N(1710)P_{11}$ contributions}
\begin{figure}[pt]
\centerline{
\epsfig{file=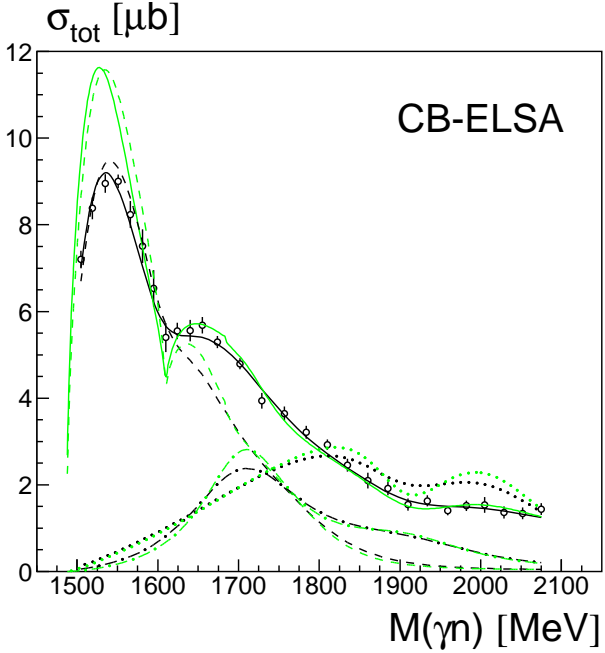,width=0.45\textwidth,clip=}}
\caption{The total cross section for the reaction $\rm\gamma n
\rightarrow \eta n$ from the deuteron target \cite{Jaegle:2008ux}.
The PWA description from Solution 2 (Paris wave function) is shown
as the solid line. The dashed line is the $S_{11}$ contribution,
dash-dotted line is $P_{11}$ contribution and dotted line is
$P_{13}$ contribution The grey (online: green) curves show the
corresponding cross sections on the free neutron (no Fermi motion)}
\label{eta_pd_tot_2}
\end{figure}

We have investigated other mechanisms for an explanation of the
bump-like structure in the region 1670 MeV. To prevent a strong
interference in the $S_{11}$ wave we forbid a direct photoproduction
of the second K-matrix pole by setting its $\gamma n$ coupling to
zero. We still observed some small interference effect in the
$S_{11}$ wave on the free neutron but it is too small to describe
the bump-like structure in the data (see dashed lines in Fig.
\ref{eta_pd_tot_2}).

 In some analyses, $N(1710)P_{11}$ has a sizable coupling to
$N\eta$, the Review of Particle Properties calculates a branching
ratio $Br(N(1710)\to N\eta)=(6.2\pm 1.0)$\%. Indeed, with suppressed
interference in the $S_{11}$ wave and absence of an exotic state,
this is the only mechanism which can explain the data. The $\chi^2$
of the fit is very similar to the solution~1.

The contributions are depicted in Fig. \ref{eta_pd_tot_2}. The
helicity couplings for the $S_{11}$ resonances calculated as
residues in the pole position are determined to
 \be
 S_{11}(1535):\; & A^n_{1/2}\! =\! -0.080 \pm 0.020\;,\;& \phi \!=\! 10^\circ \pm
 10^\circ \nonumber\\
 S_{11}(1650):\; & A^n_{1/2}\! = \! -0.020\pm 0.015\;,\;& \phi \!=\! 25^\circ \pm
 20^\circ ~~~~~~
 \label{p11_1710}
 \ee

This solution differs from the solution 1 by a different partial
wave decomposition: it has a significant contribution from $P_{11}$
in the region around 1.7 GeV which comes from the $N(1710)P_{11}$
resonance. The description of the total cross section for the Paris
wave function is shown in Fig.~\ref{eta_pd_tot_2}. This analysis
shows that there is a second possible mechanism to describe the
existing experimental data and the structure around 1.67\,GeV in
$\eta$ photoproduction.

As before, the dominant contribution stems from the $S_{11}$ wave
but in the 1700\,MeV region, the $P_{11}$ wave provides an
appreciable contribution, too. In this solution, interference
between $N(1535)S_{11}$ and $N(1650)S_{11}$ makes a visible but
small effect.

\begin{figure}[h!]
\centerline{
\epsfig{file=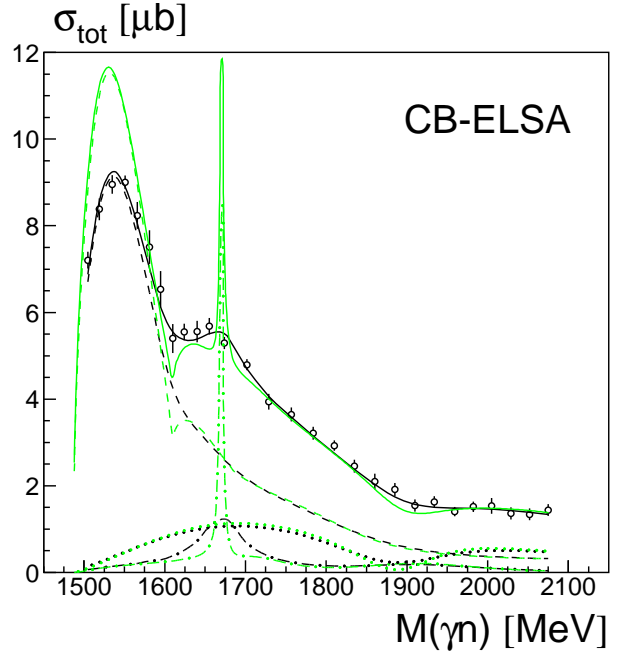,width=0.45\textwidth,clip=}}
\caption{The total cross section for the reaction $\rm\gamma n
\rightarrow \eta n$ from the deuteron target. The PWA description
from Solution 3 (Paris wave function) is shown as the black solid
line and the contributions as the colored solid lines. The dashed
curves show the corresponding cross sections on the free neutron (no
Fermi motion)} \label{eta_pd_tot_3}
\end{figure}

\subsubsection{Is there a narrow $P_{11}$ state?}

A narrow $P_{11}$ state in the region of 1670 MeV is discussed as a
candidate for the pentaquark \cite{Kuznetsov:2008ii} and is one of
the main motivations behind this analysis. In a third fit, we
followed the procedure for the solution 2 but introduced a narrow
state in the region 1670 MeV. Its mass optimized at $1670\pm
6$\,MeV. If its width was allowed to vary, the resonance became
broader and interfered with the standard $N(1710)P_{11}$ resonance.
The fits became unstable and a series of solutions were obtained in
which a relatively narrow state and the broader $N(1710)P_{11}$
interfered. Solution 3, presented in the Fig.~\ref{eta_pd_tot_3},
shows the extreme where the broad $N(1710)P_{11}$ wave is absent. In
this solution, the helicity coupling for the narrow $P_{11}$ state
is equal to 0.016\,GeV$^{-1/2}$, assuming a $\eta p$ branching to be
50\%. It is interesting to note that Azimov {\it et al.}
\cite{Azimov:2005jj,Kuznetsov:2008ii} derived a value
0.021\,GeV$^{-1/2}$ using the GRAAL data on $\eta$ photoproduction
off neutrons.

For the two $S_{11}$ resonances the following helicity couplings are
calculated:
 \be
 S_{11}(1535):\; & A^n_{1/2}\! =\! -0.076 \pm 0.015\;,\;& \phi \!=\! 25^\circ \pm
 10^\circ \nonumber\\
 S_{11}(1650):\; & A^n_{1/2}\! = \! -0.054\pm 0.015\;,\;& \phi \!=\! 20^\circ \pm
 20^\circ ~~~~~~
 \ee

\subsection{$\eta$ photoproduction on the free proton}

Finally we consider the recent conjecture of Kuznetsov et al.
\cite{Kuznetsov:2007dy} that the beam asymmetry for $\eta$
photoproduction on free protons may reveal a structure in the
1.69\,GeV region. Here, we check the compatibility of this data with
solution 1 (interference in $S_{11}$ wave) and/or with solution 3
(narrow $P_{11}$). In addition to \cite{Kuznetsov:2007dy}we include
also the beam asymmetry data for $\eta$ photoproduction from GRAAL
\cite{Bartalini:2007fg}.

\begin{figure}[h!]
\centerline{ \epsfig{file=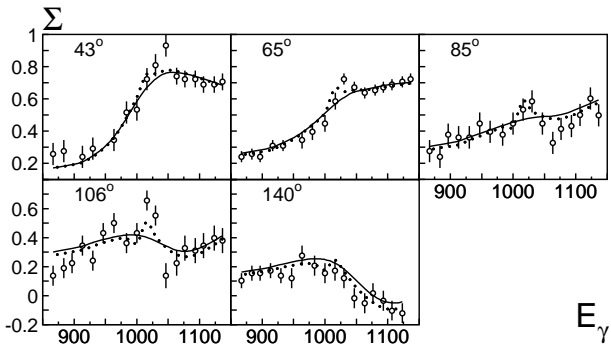,width=0.45\textwidth,clip=} }
\caption{Beam asymmetry for the reaction $\rm\gamma p \rightarrow
\eta p$ \cite{Kuznetsov:2007dy}. The PWA description is shown as
solid line (solutions 1) and dotted line (solution 3).}
\label{slava}
\end{figure}

\begin{figure}[h!]
\centerline{
\epsfig{file=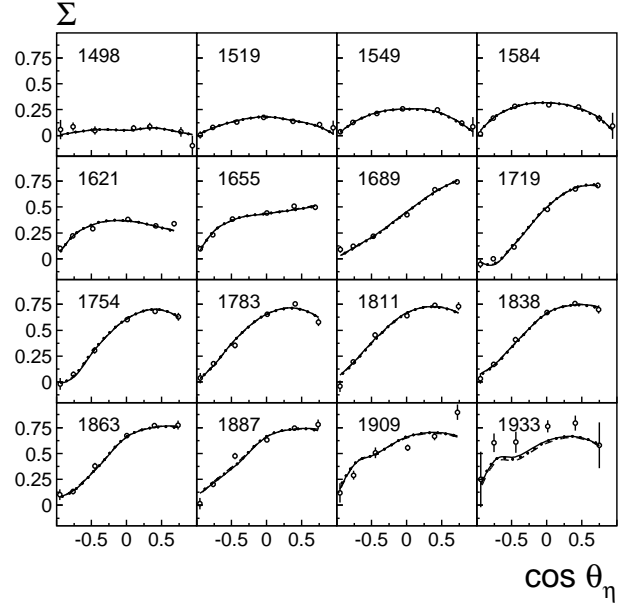,width=0.45\textwidth,clip=} }
\caption{Beam asymmetry for the reaction $\rm\gamma p \rightarrow
\eta p$ \cite{Bartalini:2007fg}. The PWA description is shown as
solid line (solutions 1) and dotted line (solution 3).}
\label{graal_gp} \vspace{-4mm}
\end{figure}

The data on $\gamma p \rightarrow p\eta$ and fit  are shown on Figs.
\ref{slava} and \ref{graal_gp}. The data are described by the
solution 1 with $\chi^2/N_{data} = 1.35$ (new data
\cite{Kuznetsov:2007dy}) and $\chi^2/N_{data} = 1.85$ (GRAAL data
\cite{Bartalini:2007fg}).
Introducing a narrow $P_{11}$ state (solution 3) results in a
$\chi^2/N_{data} = 0.95$ (new data \cite{Kuznetsov:2007dy}) and
$\chi^2/N_{data} = 1.90$ (GRAAL data \cite{Bartalini:2007fg}).
Although the solution with a narrow $P_{11}$ gives a better
description of the new data \cite{Kuznetsov:2007dy}, the fit faces
some problems. While all individual pictures in Fig. \ref{graal_gp}
exhibit a peak-like structures, they cannot be described
consistently by one resonance with one unique mass position.  Hence
new high precision data on this reactions are urgently needed if
this idea is to be pursued further.

\subsection{Helicity amplitudes}

Main parameters for the two $S_{11}$ states are given in Table
\ref{S11_table}. Pole positions and photoproduction couplings off
protons are in good agreement with a previous analysis
\cite{Thoma:2007bm}. The main change was found in the imaginary part
of the pole positions: the first pole of the $S_{11}$ amplitude
became a bit narrower and the second pole a bit broader. Both
resonances have a Flatt´e like structure, the first one due to the
$\eta N$ threshold and the second one due to $K\Lambda$. In Table
\ref{S11_table}, the position of the poles closest to the physical
region are listed. The behavior of the Flatt´e amplitude is defined
by an interplay of two poles on two sheets defined by the cut; small
instabilities in pole positions are hence not surprising.

The helicity couplings given in Table \ref{S11_table} are calculated
as the residues at the pole position and have phases. For protons,
our $N(1535)S_{11}$ helicity amplitude coincides with the PDG
estimate, for neutrons the two errors just cover the difference
\cite{Yao:2006px}. The discrepancy is due to the results reported in
\cite{Arndt:1995ak,Awaji:1981zj} while most analyses
\cite{Krusche:1995zx,Fujii:1980su,Arai:1980zj,Crawford:1980zk} quote
values which are fully compatible with our finding.

For the $N(1650)S_{11}$, we found stronger photon couplings than the
average value given in \cite{Yao:2006px}, with the differences being
at the $2\sigma$ level. We point out that our values provide a
consistent description of almost all existing data sets. Our $\gamma
n$ coupling is, for the first time, derived from $\eta$
photoproduction off neutrons (and constrained by $\pi^0$
photoproduction off neutrons).

The values in  Table \ref{S11_table} are averaged using the first
and third solution. In the solution proposing a large $P_{11}(1710)$
contribution, the $\gamma n$ coupling of the $N(1650)S_{11}$ was
found $\sim 0.4$ times smaller (see (\ref{p11_1710})).

We have not found other mechanisms to describe the bump structure in
the region 1700 MeV. An explanation given in \cite{Chiang:2002vq} as
a possible contribution from $D_{15}(1675)$ is ruled out by our
analysis. The combined fit to the $\pi$, $\eta$ photoproduction off
free protons and the deuteron and the results from the elastic $\pi
N$ scattering fixes well the branching ratio to the $\eta N$ channel
(which is $< 6$\%) and helicity couplings. Although there is a
rather large error for the $\gamma n$ coupling of the $D_{15}(1675)$
resonance, we could only reach a contribution of 0.5\,$\mu$b from
this state to the $\gamma n\to \eta n$ total cross section which is
far from the value needed for a good description of the data.

Finally, we note that we do not use different helicity amplitudes
for $\pi$ and $\eta$ photoproduction. Discrepancies, as found in the
literature for both $S_{11}$ resonances, between helicity amplitudes
derived from different data cannot occur.

\begin{table}[pb] \caption{\label{S11_table}Masses and widths (in GeV) and helicity
amplitudes of $S_{11}(1535)$ and $S_{11}(1650)$.}
\renewcommand{\arraystretch}{1.3}
 \begin{center}
\begin{tabular}{rcc}
\hline\hline
& $S_{11}(1535)$ & $S_{11}(1650)$\\
\hline \hline
Pole position (mass)& $1.505\pm0.020$ & $1.640\pm0.015$ \vspace{-1mm} \\
 (width)& $0.145\pm0.025$ & $0.165\pm0.015$  \\
PDG \qquad & $1.510\pm0.020$ & $1.655\pm0.015$ \vspace{-1mm}  \\
& $0.170\pm0.080$ & $0.165\pm0.015$  \\
\hline
$A^p_{1/2}$ (GeV$^{-1/2})$ & $0.090\pm0.025$ &  $0.100\pm0.035$ \vspace{-1mm}\\
PDG \qquad & $0.090\pm0.030$ & $0.053\pm0.016$  \\
phase & $(20\pm 15)^\circ$&  $(25\pm 20)^\circ$ \\
\hline
$A^n_{1/2}$ (GeV$^{-1/2})$ & $-0.080\pm 0.020$ & $-0.055\pm 0.020$ \vspace{-1mm} \\
PDG \qquad & $-0.046\pm0.027$ & $-0.015\pm0.021$  \\
phase & $(20\pm 20)^\circ$&  $(30\pm 25)^\circ$ \\
 \hline \hline
\end{tabular}
\end{center}
\renewcommand{\arraystretch}{1.0}
\vspace{-0.3cm}
\end{table}

In the so-called ``Single Quark Transition Model", Burkert {\it el
al.} \cite{Burkert:2002zz} extracted amplitudes for electromagnetic
transitions from proton and neutron to excited states. The
extrapolation to the photon point (read off their diagrams) are
listed in Table \ref{Models}. The agreement is excellent.

Finally we also compare our photocouplings with model calculations
\cite{KI,Li:1990qu,Bijker:1994yr,Capstick:1992uc}. In all models,
the signs are right and the magnitudes agree with the experimental
values at the 30\% level. On the basis of this data, no preference
can be given to one particular model calculation.

\begin{table}[pt] \caption{\label{Models}Model
predictions of $S_{11}(1535)$ and $S_{11}(1650)$ helicity amplitudes
for protons and neutrons (in $10^{-3}$GeV$^{-1/2})$.}
\renewcommand{\arraystretch}{1.3}
 \begin{center}
\begin{tabular}{lccccccc}
\hline\hline &&This work \hspace{-2mm}&\hspace{-2mm}\cite{Burkert:2002zz}&\cite{KI}&\cite{Li:1990qu}&\cite{Bijker:1994yr}&\cite{Capstick:1992uc} \\
$N_{1535}$\hspace{-4mm}&\hspace{-2mm}$p$\hspace{-2mm}&\hspace{-2mm}$90\pm25$\hspace{-2mm}&\hspace{-2mm}97\hspace{-2mm}&\hspace{-2mm}+147\hspace{-2mm}&\hspace{-2mm}+142\hspace{-2mm}&\hspace{-2mm}+127\hspace{-2mm}&\hspace{-2mm}+76\\
         \hspace{-4mm}&\hspace{-2mm}$n$\hspace{-2mm}&\hspace{-2mm}$-80\pm20$\hspace{-2mm}&\hspace{-2mm}-53\hspace{-2mm}&\hspace{-2mm}-119\hspace{-2mm}&\hspace{-2mm} -77\hspace{-2mm}&\hspace{-2mm}-103\hspace{-2mm}&\hspace{-2mm}-63\\
$N_{1650}$\hspace{-4mm}&\hspace{-2mm}$p$\hspace{-2mm}&\hspace{-2mm}$100\pm35$\hspace{-2mm}&\hspace{-2mm}90\hspace{-2mm}&\hspace{-2mm} +88\hspace{-2mm}&\hspace{-2mm} +78\hspace{-2mm}&\hspace{-2mm} +91\hspace{-2mm}&\hspace{-2mm}+54\\
         \hspace{-4mm}&\hspace{-2mm}$n$\hspace{-2mm}&\hspace{-2mm}$-55\pm20$\hspace{-2mm}&\hspace{-2mm}-32\hspace{-2mm}&\hspace{-2mm} -35\hspace{-2mm}&\hspace{-2mm} -47\hspace{-2mm}&\hspace{-2mm} -41\hspace{-2mm}&\hspace{-2mm}-35\\
 \hline \hline
\end{tabular}
\end{center}
\renewcommand{\arraystretch}{1.0}
\vspace{-0.3cm}
\end{table}

\section{Conclusions}

We have presented an analysis of data on photoproduction of $\eta$
(and $\pi^0$) mesons off neutrons. The analysis was motivated by a
bump structure at 1670 MeV observed in the total cross section for
$\gamma n\to n\eta$ in several experiments. There is a hot
discussion in the literature if the structure signals a resonance.
Often, it is interpreted as evidence for a pentaquark with hidden
strangeness.

We find that the data can naturally be interpreted by interference
within the $S_{11}$ wave. This is the most natural interpretation
and does not require any ad-hoc assumption. Other interpretations
can, however, not be ruled out. The $N(1650)S_{11}$ may have a small
coupling to $n\gamma$. Then, the $P_{11}$ amplitude plays a more
significant role. For an appropriate choice of parameters, a narrow
$P_{11}$ can be introduced and the data are well described. Hence
the data do not support the need to introduce a narrow resonance
but, for a suited set of parameters, the existence of a narrow
resonance is also not ruled out. Fluctuations in recent beam
asymmetry data for $\gamma p\to p\eta$ may serve as an indication
for a narrow structure at 1670\,MeV but fits without it provide a
reasonable description of the data as well.

A second aspect of the data is the determination of helicity
amplitudes. Our values are mostly consistent with those listed by
the Particle Data Group. Comparison with model calculations show
reasonable agreement but none of the models gives strikingly better
results than the other models. Our values agree very well with a fit
to electroproduction data using the ``Single Quark Transition
Model", Burkert {\it el al.} \cite{Burkert:2002zz}.

\section*{Acknowledgements}
We would like to thank the CBELSA collaboration for many useful
discussion and the GRAAL collaboration for allowing us to use their
data prior to publication. We acknowledge financial support from the
Deutsche Forschungsgemeinschaft (DFG-TR16) and the Schweizerische
Nationalfond. The collaboration with St. Petersburg received funds
from DFG and RFBR.

\end{document}